# Emergency Locator Transmitters in the Era of More Electric Aircraft: A Comprehensive Review of Energy, Integration and Safety Challenges

Juana M. Martínez-Heredia [1,*], Adrián Portos [1], Marcel Štěpánek [2], and Francisco Colodro [1]

[1]     Department of Electronic Engineering, Universidad de Sevilla, Seville, 41092, Spain
[2]     Department of Aviation Technology, University of Defence, Brno, Czech Republic
*      Correspondence: jmmh@us.es

**Abstract**

The progressive electrification of aircraft systems under the more electric aircraft (MEA) paradigm is reshaping the design and qualification constraints of safety-critical avionics. Emergency locator transmitters (ELTs), which are essential for post-accident localization and search and rescue (SAR) operations, have evolved from legacy 121.5/243 MHz beacons to digitally encoded 406 MHz systems, typically retaining 121.5 MHz as a homing signal in combined units. In parallel, the modernization of the Cospas-Sarsat infrastructure, especially MEOSAR, together with multi-constellation global navigation satellite system (GNSS) integration and second-generation beacon capabilities, is reducing detection latency and enabling richer distress messaging. However, MEA platforms impose stricter constraints on available power, thermal management, wiring density, and electromagnetic compatibility (EMC). As a result, ELT performance increasingly depends not only on the device itself, but also on its installation conditions and on the aircraft's overall electrical environment. This review summarizes the ELT architectures and activation/operational cycles, outlines key technological milestones, and consolidates the main integration challenges for MEA, with emphasis on energy autonomy, battery qualification frameworks, EMC and installation practices, and survivability-driven failure modes (e.g., antenna/feedline damage, mounting, and post-impact shielding). Finally, emerging trends include ELT for distress tracking (DT), energy-based designs, advanced health monitoring, and certification-ready pathways for next-generation SAR services are discussed, highlighting research directions that can deliver demonstrable, certifiable gains in reliability, energy efficiency, and robust integration for future electrified aircraft.

## 1. Introduction

The progressive electrification of aircraft systems is driving a profound transformation in modern aviation. Under the more electric aircraft (MEA) paradigm, hydraulic, pneumatic, and mechanical subsystems are increasingly being replaced by electrically powered alternatives, enabling improvements in energy efficiency, weight reduction, system controllability, and overall aircraft performance [1]. This transition, however, also introduces new challenges related to the design, integration, and certification of safety-critical onboard equipment, particularly those systems that must remain operational under extreme conditions or after catastrophic events.

Among such safety-critical systems, emergency locator transmitters (ELTs) play a fundamental role in post-crash survivability by enabling the rapid localization of downed aircraft and significantly improving the effectiveness of search and rescue (SAR) operations [2]. ELTs constitute a mandatory component in most civil aviation platforms and represent the final technological link between an aircraft accident and the initiation of coordinated rescue missions.

Since their introduction more than four decades ago, ELTs have evolved in response to regulatory updates, technological advances in radiofrequency communications, and the progressive modernization of the global satellite-based SAR infrastructure [3]. These developments have led to substantial improvements in detection capability, localization accuracy, and global coverage.

At the same time, the increasing electrification of aircraft systems has altered the operational and electrical environment in which ELTs must function, raising new questions regarding their integration, robustness, and long-term reliability. Traditional ELT designs have primarily focused on ensuring reliable activation, sufficient radiated power, and guaranteed autonomous operation during distress situations. However, the transition toward MEA platforms introduces additional constraints associated with more complex electrical architectures, increased power density, and tighter integration with avionics systems [1-5]. As a result, ELTs are now required to operate within environments characterized by more demanding electrical interfaces, stricter electromagnetic compatibility (EMC) requirements, and more constrained installation conditions, while continuing to meet stringent safety and certification standards.

Despite the critical role of ELTs in aviation safety, the existing literature lacks a comprehensive review that analyzes their technological evolution and integration challenges from the perspective of the MEA paradigm. Previous surveys have often focused on regulatory aspects or satellite-based detection performance, but they rarely examine ELTs from an electrical and energy-oriented viewpoint or assess how next-generation aircraft architectures impact their design, integration, and certification.

This article addresses this gap by presenting a comprehensive review of the state of the art in emergency locator transmitters technology, with particular emphasis on aspects that are critical for more electric aircraft. The main contributions of this review are as follows:

(i) a structured overview of ELT fundamentals, architectures, and operational principles;

(ii) an analysis of the technological evolution of ELTs, highlighting key advances in signaling, localization, and activation mechanisms;

(iii) a detailed discussion of electrical and energy-related challenges associated with integrating ELTs into MEA power architectures; and

(iv) an assessment of emerging trends and research opportunities aimed at improving energy efficiency, reliability, and system integration in next-generation ELTs.

The remainder of this paper is organized as follows. Section 2 introduces the fundamentals of emergency locator transmitters, including their architectures, classifications, and operating principles. Section 3 reviews the technological evolution of ELTs and the main advances achieved to date. Section 4 analyzes the electrical and energy aspects that are particularly relevant for ELT integration in more electric aircraft. Section 5 discusses certification and safety considerations. Section 6 outlines emerging trends and future research opportunities. Finally, Section 7 presents the main conclusions of the study.

## 2. Fundamentals of Emergency Locator Transmitters

Emergency locator transmitters (ELTs) are autonomous radio beacons designed to support search and rescue (SAR) operations by transmitting distress signals following an aircraft accident. Their primary purpose is to enable rapid detection and localization of downed aircraft, thereby reducing rescue response times and increasing post-crash survivability. ELTs are typically battery-powered and are intended to remain operational even when the aircraft electrical system is unavailable or severely damaged. ELTs constitute a mandatory safety component for most civil aircraft categories under international aviation regulations [2] and operate as part of the global satellite-based SAR infrastructure coordinated by the Cospas–Sarsat programme [3].

ELTs belong to a broader family of distress alerting devices operating at 406 MHz and are supported by the Cospas-Sarsat system in different operational domains:

- ELTs are used in the aeronautical domain;

- Emergency position-indicating radio beacons (EPIRBs) are used primarily in the maritime domain. Initially, like ELTs, they only transmitted at 121.5 MHz, but today their signal is detected by the Cospas-Sarsat satellites only at 406 MHz. They are similar to ELTs, with the only differences being some characteristics necessary for operating over water. The device must float (most ELTs do not require this as they are rigidly attached to the aircraft) and must have a powerful light that allows rescue services to locate the accident site at night or in low visibility conditions.

- Personal locator beacons (PLBs) are carried by individuals who engage in outdoor activities such as hiking, climbing, or recreational boating. They are useful in areas where there is no mobile coverage and where it is necessary to send an emergency alert. Initially, their use was more restricted, but it expanded when governments allowed civilians to access the Cospas-Sarsat network via personal beacons between the 1990s and 2000s. Some of the main differences compared to ELTs and EPIRBs are their smaller size, which improves portability, and the fact that there are no regulations mandating their use.

- Ship Security Alert System (SSAS), unlike ELTs, focuses on security threats rather than location following an accident. The SSAS is a mandatory system on board commercial vessels that allows a silent and discreet alert to be sent to the authorities if the vessel is under a security threat, such as piracy, terrorism, or unauthorized boarding.

This broader context is useful to distinguish ELTs as safety-driven, crash-related systems whose design is tightly coupled to SAR infrastructure and aeronautical certification requirements.

ELTs are specifically engineered to function independently of the aircraft's main power systems and avionics following a crash event. To this end, they incorporate dedicated activation mechanisms, autonomous power supplies, and robust radiofrequency transmission systems capable of operating under extreme mechanical, thermal, and environmental conditions.

*2.1 Operational Principles and Functional Architecture*

From a functional perspective, a conventional ELT [4,5] can be decomposed into some main subsystems, which can be observed in Figure 1. The autonomous power supply, typically based on primary lithium batteries, provides the required operational endurance independent of aircraft power availability. The sensing and activation unit is responsible for detecting abnormal conditions associated with a crash or ditching event. Activation may be automatic, manual, or remotely commanded from the cockpit. The control and processing unit manages activation logic, message formatting, timing of transmissions, and, when applicable, integration of positioning data. The radiofrequency (RF) transmission module is responsible for generating and radiating the distress signal at the standardized frequencies defined by international SAR regulations. Finally, the antenna system is designed to ensure adequate radiation characteristics under uncertain post-crash orientations and environmental conditions.

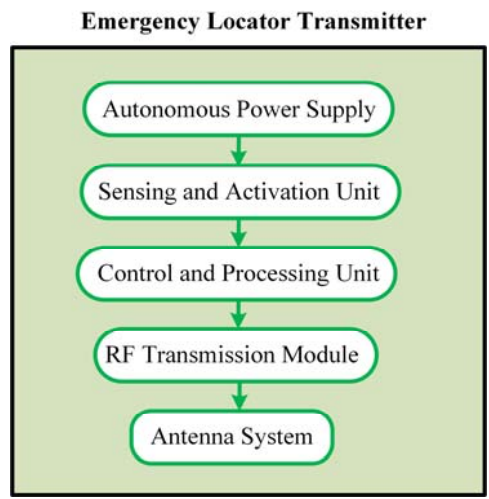

**Figure 1.** System architecture for a conventional ELT.

Once activated, the ELT transmits periodic distress messages according to predefined protocols, enabling detection by satellites, airborne receivers, or ground-based systems depending on the operating frequency.

*2.2 Classification of Emergency Locator Transmitters*

International aviation authorities, including International Civil Aviation Organization (ICAO), Federal Aviation Administration (FAA), and European Union Aviation Safety Agency (EASA), classify ELTs mainly according to their installation and deployment characteristics [2], [4,5]. Table 1 summarizes the commonly used ELT categories, including fixed, portable, and deployable variants, as well as survival-oriented configurations.

Each category is subject to specific performance, environmental, and certification requirements defined in applicable Technical Standard Orders (TSOs) and European Technical Standard Orders (ETSOs).

**Table 1.** ELT classification (summary).

| Type | Name | Typical characteristics and intended use |
|---|---|---|
| ELT(AF) | Automatic Fixed | Permanently installed in the aircraft; automatically activates upon impact and can also be activated remotely from the cockpit. |
| ELT(AP) | Automatic Portable | Installed on the aircraft but removable; can activate automatically and can be carried/operated manually after the accident. |
| ELT(AD) | Automatic Deployable | Designed to automatically detach/eject (often for overwater operations); may float and improve survivability of the transmitting unit and antenna orientation. |
| ELT(P) | Portable | Portable beacon intended to be carried and activated manually (e.g., for survival scenarios). |
| ELT(W)/ ELT(S) | Water-activated / Survival | Configurations intended for ditching/overwater and/or survival kit use, typically emphasizing manual handling and endurance in harsh environments. |

ELT installation location is a critical practical aspect because it influences both survivability and radio propagation after a crash. A common practice is to place the ELT toward the aft fuselage/tail section, since this area may have a higher probability of remaining structurally intact in certain crash scenarios and may offer improved antenna visibility depending on wreckage orientation. The installation must also account for antenna placement, coaxial routing robustness, and the likelihood of post-impact shielding by metallic/composite structures, since these factors strongly affect the probability of successful transmission during the SAR response [6].

*2.3 Operating Frequencies and Signaling Protocols*

Historically, ELTs operated on analog distress frequencies of 121.5 MHz (civil aviation) and 243 MHz (military aviation), supporting local homing by radio direction finding. These legacy beacons typically transmitted an AM carrier modulated with a swept-tone (warbling) audio signal, i.e., a tone whose frequency is periodically swept over a predefined range, producing a distinctive siren-like signature. This modulation improves practical detectability and facilitates the final homing phase once the SAR assets are in the vicinity of the accident site. However, due to limitations in detection reliability and high false alarm rates, satellite monitoring of these frequencies was discontinued by the Cospas–Sarsat system in 2009 [3,7].

Modern ELTs predominantly rely on 406 MHz digital distress transmissions aligned with the international Cospas–Sarsat framework [3], [5], allowing beacon identification, reduced false alert rates, and improved global detection, often including GNSS-encoded position for improved localization accuracy. In many contemporary installations, 121.5 MHz is retained as an auxiliary homing signal in combined 406/121.5 MHz ELTs, supporting short-range direction finding during the final localization stage after the initial satellite alert.

*2.4 Interaction with the Cospas–Sarsat Search and Rescue System*

ELTs are designed to interface with the Cospas-Sarsat international satellite-aided SAR system, which detects 406 MHz distress signals and routes alerts to rescue coordination centers. The Cospas-Sarsat system [3,8] is composed of:

• distress radio beacons (ELTs for aviation use, EPIRBs for maritime use, and PLBs for personal use) that transmit signals during distress situations;

• instruments on board satellites in geostationary, medium-altitude Earth orbits, and low-altitude Earth orbits, which detect the signals transmitted by distress radio beacons;

• ground receiving stations, referred to as local users terminals (LUTs), which receive and process the satellite downlink signals to generate distress alerts; and

• mission control centers (MCCs) which receive alerts produced by LUTs and forward them to rescue coordination centers (RCCs), search and rescue points of contacts (SPOCs), or other MCCs.

Figure 2 illustrates the end-to-end interaction with the Cospas–Sarsat system.

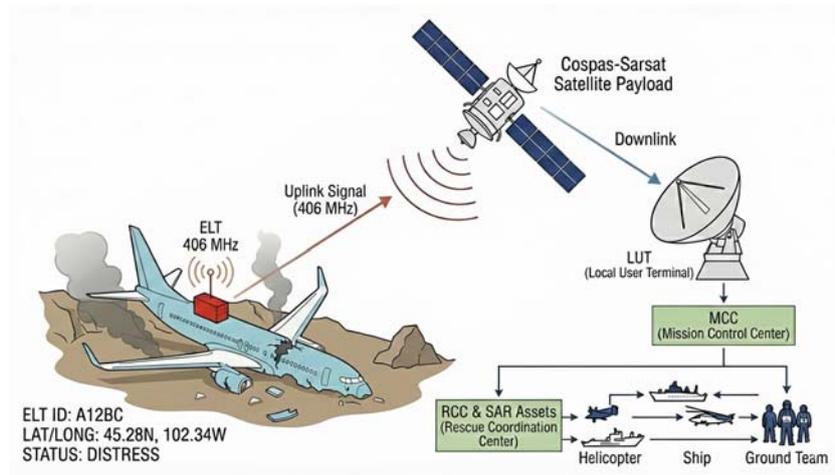

**Figure 2.** Interaction of an ELT with the Cospas–Sarsat system.

The system is supported by three complementary satellite segments:

- LEOSAR (Low Earth Orbit SAR): The satellites are located at a low altitude, approximately between 700 km and 1,000 km above the Earth's surface. They orbit from pole to pole and take about 100 minutes to circle the Earth. This segment provides global coverage with Doppler-based localization capability. Because LEO satellites move relative to the beacon, they can support location estimation even when the GNSS coordinates are not embedded, although detection may depend on the satellite pass timing.

- GEOSAR (Geostationary SAR): The satellites are placed in geostationary orbits at an altitude of about 35,786 km above the equator. This segment offers near-continuous regional coverage and potentially faster detection because GEO satellites remain fixed relative to the Earth. However, geostationary satellites do not provide Doppler-based location; therefore, location performance benefits strongly from GNSS-encoded distress messages.

- MEOSAR (Medium Earth Orbit SAR): The satellites orbit at an intermediate altitude, generally between 19,000 km and 24,000 km. This segment combines wider coverage with improved detection latency and enhanced localization performance through multi-satellite reception geometry (e.g., time-difference processing), generally improving robustness and timeliness compared with legacy-only architectures.

The evolution toward MEOSAR has significantly enhanced SAR performance, influencing both the operational characteristics and design requirements of modern ELTs.

*2.5 Activation Modes and Operational Cycle*

ELTs may be activated through multiple mechanisms:

- Automatic activation, typically based on mechanical or inertial sensors that respond to impact or deceleration thresholds.
- Manual activation, either remotely via a cockpit control panel or directly by crew/passengers on the ELT unit.
- Water-activated deployment, particularly for ELT-AD configurations intended for overwater operations.

Following activation, the ELT enters a predefined operational cycle in which distress messages are transmitted at regular intervals until battery depletion, manual deactivation, or rescue termination. These operational features are foundational for later discussions of energy endurance, false activation mitigation, and integration within more complex aircraft electrical environments.

*2.6 Power Supply and Endurance Requirements*

A defining characteristic of ELTs is the use of an independent power source, usually a dedicated battery pack, designed to ensure continued operation during and after an emergency event, even if the aircraft electrical system becomes unavailable. Certification standards generally require a minimum autonomous operating time of 24 to 48 hours, depending on the ELT category and the operating temperature range [4, 5]. These endurance constraints directly influence the overall ELT design, including transmission duty cycle, internal power management, and the selection of components with low standby and operational consumption.

In current certified implementations, primary lithium battery technologies are widely adopted because they provide high energy density, long shelf life, and acceptable performance over wide temperature ranges. However, the power subsystem is not defined solely by nominal capacity: battery performance, replacement intervals, and environmental robustness (e.g., vibration, shock, humidity, and thermal exposure) are tightly regulated within the ELT certification framework. Consequently, endurance requirements drive key engineering decisions such as battery chemistry and packaging, protection circuitry, thermal robustness, and the implementation of periodic self-test and monitoring functions to ensure readiness throughout the service life.

Although these aspects are common to all aircraft, they become particularly relevant when considering ELTs within the broader context of highly electrified platforms, where installation constraints and the surrounding thermal/electromagnetic environment may be more demanding [1], [9]. Accordingly, Section 4.1 revisits ELT power supply and autonomy from an MEA-oriented perspective, focusing on the electrical and energy implications of meeting endurance requirements under more stringent integration constraints.

## 3. Evolution of ELT Technologies

The development of emergency locator transmitters has been shaped by two converging drivers: (i) the need to reduce false alerts and improve real-world reliability after accidents, and (ii) the continuous evolution of the global SAR infrastructure supporting detection and localization. Over the last decades, ELTs have transitioned from purely analog beacons aimed at local homing to digitally encoded systems designed for global satellite detection, identification, and, increasingly, precise position reporting [3,6,10].

*3.1. Historical Transition from 121.5 MHz Alerting to 406 MHz Digital Distress*

Operational experience and policy decisions progressively consolidated 406 MHz as the primary channel for satellite distress alerting [11,12], while 121.5 MHz remained mainly relevant for close-in homing (particularly in combined 406/121.5 MHz ELTs) [3,13]. The operational implications of this transition have been quantified using SAR mission databases, showing statistically significant improvements in search duration for aircraft equipped with 406 MHz ELTs relative to legacy 121.5 MHz units in general aviation contexts [10].

A key milestone in this transition was the cessation of satellite processing of 121.5 MHz and 243.0 MHz beacon signals on 1 February 2009, after which Cospas–Sarsat satellite processing focused on 406 MHz alerts [7].

From a maintenance and continued-airworthiness standpoint, the "406 MHz era" introduced practical changes in verification: reliable functional checks typically require dedicated 406 MHz test equipment and procedures that go beyond legacy "AM radio checks" traditionally associated with 121.5 MHz-only ELTs [14].

*3.2. Standardization, Type Approval, and Interoperability Requirements*

Complementing the technical specification, type approval processes (and related documentation) have become a critical part of ensuring consistent performance, reproducibility, and compatibility across manufacturers and implementations. Although the full type-approval standard is not always distributed as a single open PDF, it is repeatedly referenced as the baseline for verifying compliance in Cospas–Sarsat type-approval pathways [15].

The shift from analog tones to digital distress messaging required a robust interoperability layer across the full distress-alerting chain. Within the Cospas–Sarsat system, this is realized through technical specifications defining beacon transmission characteristics, coding, and message structure across beacon classes (ELT/EPIRB/PLB) [16]. Complementarily, type approval pathways and associated documentation help ensure repeatability and compatibility across manufacturers [17]. In parallel, aviation-specific minimum operational performance standards (MOPS) and authority approval documents (e.g., RTCA MOPS and TSO/ETSO frameworks) translate system-level requirements into certifiable equipment behavior and testable performance [4,5,13,16].

Recent ETSO baselines explicitly reference EUROCAE ED-62B (and amendments) as the minimum performance standard for 406 MHz ELTs, reflecting the maturation of requirements for GNSS position information, crash safety, antenna/cabling, and distress-tracking-related functions [17].

At the equipment level, these frameworks impose constraints on modulation and coding, burst timing and repetition behavior, and message-field definitions, which are features that directly affect detection probability, false-alert handling, and end-to-end interoperability across the space and ground segments [13,16]. Complementary analyses in the open literature have characterized spectral properties and signal behavior of 406 MHz transmissions, providing technical underpinning for receiver design and robustness discussions [18].

*3.3. Position Reporting and Dual-Frequency Operation (406 MHz and 121.5 MHz)*

A major evolutionary step has been the integration of GNSS-derived location into the 406 MHz distress message (where supported), which reduces uncertainty and can accelerate SAR decision-making and deployment [3,16]. Encoded position is particularly valuable in scenarios where the satellite segment can rapidly detect an activation but cannot independently compute a location without position data embedded in the message.

Importantly, many aviation ELTs remain dual-frequency: 406 MHz provides global alerting and identification (often including encoded position), while 121.5 MHz is retained as a homing signal to support final localization by nearby SAR aircraft or ground teams once responders are operating within the approximate search area [3,13].

*3.4. Satellite Segment Evolution (LEOSAR, GEOSAR, and MEOSAR)*

ELT capabilities cannot be separated from the satellite infrastructure that detects and locates beacon bursts. The Cospas–Sarsat space segment historically included LEOSAR and GEOSAR, each with well-known tradeoffs (e.g., revisit time versus near-continuous coverage). The introduction of MEOSAR, leveraging many medium-earth-orbit navigation satellites, was designed to reduce detection latency and improve coverage redundancy compared with legacy architectures [19].

In LEOSAR systems, Doppler-based localization may produce two possible position estimates, one of which is a mirror solution. This ambiguity is resolved using additional observations. The operational message flow and the procedures used to resolve this ambiguity are described in the Cospas–Sarsat handbooks [8]

From an ELT perspective, MEOSAR evolution reinforces the importance of stable burst characteristics and consistent message integrity, since multi-satellite processing benefits from predictable, standard-compliant transmission behavior and correctly encoded payload fields [13,16,19].

*3.5. System-Level Evolution: Second-Generation Beacons, Return Link Service, and Distress Tracking*

Recent developments extend the traditional post-crash beacon paradigm toward system-level capabilities. Second-Generation Beacons (SGB) introduce updated signaling and transmission behaviors intended to improve detection and capacity (particularly in MEOSAR-oriented processing) while remaining grounded in Cospas–Sarsat specifications [20]. In parallel, service enhancements such as Return Link Service (RLS) and Distress Tracking ELTs (ELT(DT)) expand the operational concept beyond post-impact alerting by adding acknowledgment and/or higher-rate tracking modes in distress scenarios. These capabilities are reflected in the evolving beacon specification baseline, including dedicated provisions for RLS and ELT(DT) features (e.g., specific modes of operation and cancellation messaging) [16].

To consolidate the main steps in ELT evolution and highlight their implications for integration in electrically dense aircraft, Table 2 summarizes key milestones alongside the enabling SAR-system changes and the corresponding standardization baseline.

**Table 2.** Key milestones in ELT evolution and relevance to more-electric aircraft (MEA), including enabling SAR-system changes and associated standardization baselines.

| Period / Year | Milestone | SAR interaction change (Cospas–Sarsat) | Relevance to MEA (energy / integration / safety) | Key standard / document |
| --- | --- | --- | --- | --- |
| | | | | |
| 1970s | Early civil aviation ELTs at 121.5/243 MHz | Primarily aural detection; limited satellite-oriented processing and no unique identification. | Baseline "standalone" designs; limited functionality motivates later data/position integration. | [2]; [3] |
| 1979–1985 | Cospas–Sarsat established (first satellite 1982; operational 1985) | Global satellite-aided distress alerting and distribution chain defined (space + ground + MCC→ RCC). | System-level survivability expectations strengthen battery autonomy and crash survivability requirements. | [3] |
| 1998 | GEOSAR introduced to complement LEOSAR | Near-continuous regional detection; GEO cannot Doppler-locate → encoded position becomes critical. | Drives GNSS/position-encoding adoption and tighter RF/antenna and EMC integration. | [19]; [16] |
| 1 Feb 2009 | End of satellite processing for 121.5/243 MHz | Satellite processing transitions to 406 MHz-only alerting, reinforcing the 406 MHz ecosystem. | Accelerates retrofit to 406 MHz ELTs and associated installation, GNSS interfacing, and qualification. | [10]; [13] |
| 2010–2012 | Qualification and approval baselines consolidated (MOPS/TSO/ETSO) | Standard-compliant message/burst behavior enables consistent end-to-end processing across segments. | MEA electrical density increases EMI risk → qualification and power-quality constraints become central. | [13]; [4]; [5] |
| 2016→ | MEOSAR operational roll-out via GNSS constellations | More satellites in view →faster detection and improved location compared with LEO/GEO alone. | Supports higher-performance modes (tracking, quicker alerting) but increases dependence on integration (data, antennas, power). | [19] |
| 2016–2021 | Second-Generation Beacons (SGB) specified | Updated signaling and transmission behaviors optimized for MEOSAR detection/capacity/robustness. | Impacts power budgeting, RF/antenna integration, and EMC margins due to advanced signaling. | [17] |
| 2020–2021 | Beacon specification expanded for modern services | Dedicated provisions for encoded position, RLS-related behavior, and ELT(DT)-specific messaging. | Increases functional expectations and verification scope, affecting interfaces and system integration. | [16]; [15] |
| 2020–2025 | RLS and Distress Tracking adoption pathways | Acknowledgment and/or higher-rate tracking modes expand beyond post-impact alerting in distress scenarios. | Major MEA coupling: aircraft-power input + battery fallback, avionics trigger interfaces, and expanded verification scope. | [16]; [2] |

Table 2 provides a compact reference for the evolution pathway discussed above, and it is used in the next section to motivate the energy, interface, and EMC constraints that become more stringent under the MEA paradigm.

*3.6. Representative Beacon Design Studies and Implementation Lessons*

Selected beacon design studies bridge specifications and implementation. ELT signaling oriented to very large scale integration (VLSI) at 406 MHz shows how standardized burst formats map to digital pipelines, clocking, and energy per bit, with implications for the linearity and verification [21]. Enhanced ELT developments for general aviation document architecture choices (activation logic, encoding, RF chain) and test strategies that inform practical considerations of compliance and survival [22]. Unmanned aerial vehicle (UAV) beacon implementations demonstrate how compact platforms manage GNSS acquisition, RF-burst energy, and duty-cycling under tight "size, weight, and power" (SWaP) constraints, providing transferable lessons for ELT(DT) in MEA environments [23]; long-range, low-cost variants further show how range and affordability can be balanced in compact form factors [24]. Together, these studies complement the Cospas–Sarsat and MOPS/TSO/ETSO baselines by exposing the design artifacts and measurements that underpin certifiable behavior [15–17].

*3.6. Summary and Implications for MEA*

In summary, ELT evolution reflects: (i) the transition from analog homing to 406 MHz digital distress alerting with unique identification and standardized messaging; (ii) growing reliance on formal specifications and verification pathways to ensure interoperability; (iii) improved technical robustness including GNSS-enabled position reporting; and (iv) modernization of the SAR space segment toward MEOSAR. Collectively, these developments define the functional and regulatory baseline for the energy, interface, and EMC challenges associated with integrating ELTs in power-dense MEA electrical architectures, addressed in Section 4 [1,9].

## 4. Electrical and Energy Aspects for ELT Integration in More Electric Aircraft

The MEA paradigm reshapes the electrical environment in which safety-critical systems operate by increasing onboard electrical power density, the number of power-electronic converters, and the complexity of distribution architectures [25-28]. Within this environment, ELTs must preserve reliable activation and distress transmission not only in nominal conditions but also during abnormal electrical events and post-accident scenarios.

This section focuses on electrical and energy aspects that are most relevant for integrating ELT into MEA platforms: (i) power-supply modes and operating states, (ii) energy budgeting and endurance, (iii) robustness to power-quality disturbances, (iv) EMC/EMI considerations from an integration viewpoint, and (v) installation constraints affecting antennas and wiring. The discussion is anchored to ELT approval baselines and Cospas–Sarsat specifications, together with MEA power-system and EMC studies that explicitly address onboard microgrids and evolving electromagnetic environments.

Figure 3 provides a system-level view of the coupled electrical, EMC, installation, and survivability factors that condition ELT/ELT(DT) performance in MEA platforms. It is intended as a roadmap for Sections 4.1–4.6, highlighting the interactions between power sourcing and energy storage, converter-dominated electrical environments, EMC/EMI constraints, and antenna/cabling installation robustness.

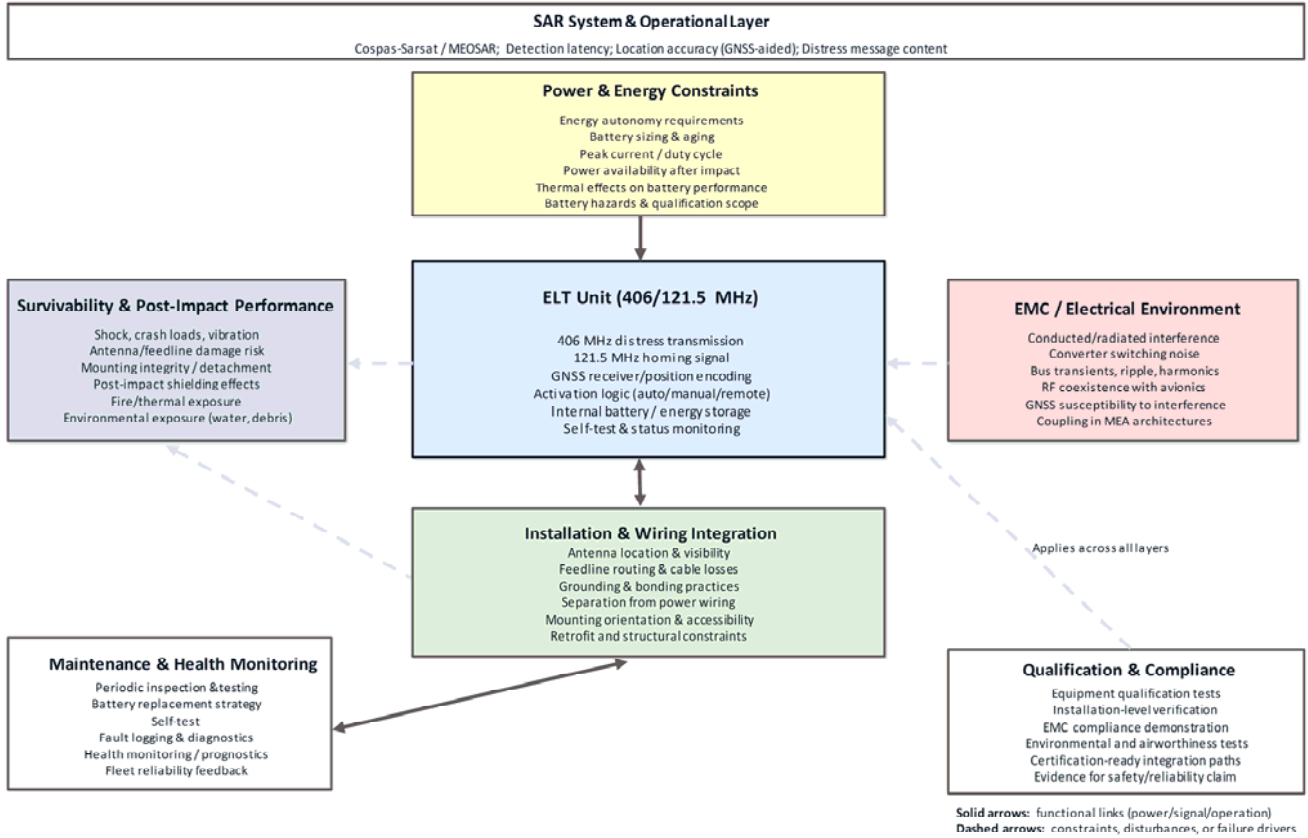

**Figure 3.** System-level view of ELT integration in MEA, highlighting the coupled interactions among power/energy sourcing and endurance, converter-dominated electrical environments, EMC/EMI constraints, antenna/cabling installation robustness, and post-impact survivability. The diagram serves as a roadmap for the electrical and energy aspects discussed in Sections 4.1–4.6.

*4.1. Criticality and Power Supply Requirements for ELT/ELT(DT) in MEA*

MEA power systems increasingly resemble isolated onboard microgrids in which multiple AC/DC voltage levels coexist and are coupled through power electronic converters; system reliability and load-management strategies therefore become central design drivers [25]. In this context, ELTs and ELT(DT) functions should be treated as safety-critical loads whose availability must be preserved under degraded electrical conditions such as generator faults, bus reconfigurations, or intentional load shedding [27] consistently with reliability-oriented DC-architecture synthesis and dynamic-stability analyses in more-electric aircraft [28].

From a functional standpoint, ELT(DT) operation modifies the traditional view of power supply: during normal flight, the unit may be powered from the aircraft electrical system to support autonomous distress tracking, but it must switch to internal battery operation after loss of aircraft power without compromising its Global Aeronautical Distress and Safety System (GADSS)-mandated distress function [29,30]. Cospas–Sarsat material on ELT(DT) and ICAO guidance on GADSS [25] explicitly emphasize continued operation after power loss, reinforcing the classification of ELT(DT) as a high-priority load within onboard power-management hierarchies.

*4.2. Electrical Integration in MEA Power Architectures (Distribution, Conversion, and Stability)*

In both more electric aircrafts (MEA) and all electric aircrafts (AEA), distribution-related choices (e.g., zonal architectures, DC buses) and conversion choices (multi-stage conversion chains, converter-dominated loads) directly influence power quality, fault tolerance, and the ability to support critical loads. Reliability-driven design of DC power distribution architectures for MEA highlights that survivability constraints and topology selection are decisive for ensuring that essential loads remain supplied following failures or reconfigurations [25-27].

From the standpoint of system dynamics, converter-dominated loads can exhibit constant-power-load (CPL) characteristics that reduce stability margins and can trigger oscillatory behavior on DC buses. Recent analyses show that CPL effects can lead to voltage/current oscillations beyond power-quality limits, emphasizing the need for coordinated design of converters, bus impedance, and load-management policies [28].

*4.3. Power Sourcing Strategies and Energy Storage Sizing for ELT/ELT(DT)*

ELTs are historically designed for functional independence from aircraft power through dedicated internal batteries, ensuring operation after catastrophic events [4,5]. However, ELT(DT) capabilities introduce higher transmission rates and in-flight operation, which tighten energy-budget constraints and motivate a dual sourcing concept: initial aircraft power (when available) and guaranteed battery endurance after disconnection [29,30]. As summarized in Figure 3, ELT/ELT(DT) energy autonomy in MEA depends not only on battery sizing but also on the switchover behavior between aircraft power and internal energy storage under degraded conditions.

Development work on enhanced ELTs for general aviation has already explored architectures that combine aircraft power with rechargeable energy storage, aiming to improve endurance and maintain beacon availability under a wide range of scenarios. Operational guidance for ELT(DT) within the GADSS framework indicates rapid initiation of distress transmissions on a seconds-scale, increased burst repetition during initial distress phases, and endurance expectations for battery-backed distress tracking of several hours after loss of aircraft power, with extended operation when combined with post-crash ELT functions. These duty-cycle changes, together with the environmental and survivability constraints identified in full-scale ELT system tests, must be reflected in battery sizing, cell technology selection, thermal design, and verification planning for MEA installations [16,18,29-38].

*4.4. Electromagnetic Compatibility in MEA: Standards and Design-Stage Mitigation*

MEA electrification increases exposure to conducted and radiated interference due to dense harnessing, high-power switching converters, and tighter physical integration. For airborne equipment, environmental and EMC qualification commonly relies on RTCA DO-160G test procedures [39], which are identified by FAA AC 21-16G as an acceptable means of demonstrating environmental qualification for compliance purposes [40]. The European counterpart (EUROCAE ED-14G) is technically aligned and is often referenced as the equivalent qualification baseline for the same test philosophy. Figure 3 also highlights that EMC robustness must be addressed as a system-integration problem, since converter-dominated MEA architectures can affect both ELT transmission and GNSS-assisted positioning.

Beyond qualification, MEA design increasingly benefits from EMC assessment during early design stages. Device-level simulation and conducted-EMI analysis methods have been proposed for MEA microgrids to predict conducted emissions and susceptibility impacts before hardware integration, supporting a design-for-EMC workflow rather than late-stage troubleshooting [41]. Complementarily, EMC-focused studies of all-electric aircraft highlight that the onboard electromagnetic environment is evolving due to higher-power electric powertrains and pervasive conversion stages, potentially stressing existing standards and installation practices [42].

*4.5. Integration Constraints: Antennas, Cabling, and Installation Robustness*

The ELT antenna system and its feeder/cabling form a safety-critical path: even if the transmitter meets its performance specification, effectiveness can be compromised by antenna damage, connector disengagement, or post-crash shadowing. Full-scale general-aviation crash tests and system-performance analyses have shown that installation details (particularly ELT-to-antenna cabling, connector integrity, and mounting) can be decisive in determining whether a distress signal is successfully radiated after impact, motivating enhanced installation guidance and continued-airworthiness oversight [32, 6]. As shown in Figure 3, antenna/feedline routing and installation integrity form a critical path linking electrical integration constraints with post-impact ELT effectiveness.

For MEA platforms, the higher wiring density and routing constraints associated with complex electrical architectures can increase coupling and reduce placement flexibility, making installation-level mitigations (routing, bonding/grounding, shielding strategy, connector retention) especially important to preserve both 406 MHz distress transmission and any GNSS-derived position used for encoded location. For ELT(DT), the need to deliver timely GNSS position and state information further strengthens the case for treating antenna/coax routing and GNSS interfacing as a safety-critical path in installation documentation and zonal-safety assessments [43-45].

*4.6. Battery Qualification and Hazards*

Energy storage for safety-critical functions introduces hazards that must be managed alongside endurance and survivability considerations. For rechargeable lithium battery systems used in aircraft installations, minimum performance and safety requirements are defined in RTCA DO-311A [33]. EASA ETSO-C179b [34] explicitly references DO-311A as the applicable standard for rechargeable lithium cells, batteries, and battery systems intended to power aircraft equipment, including emergency systems.

For non-rechargeable lithium batteries, RTCA DO-227A [35] provides the minimum operational performance standard, and EASA ETSO-C142b [36] references DO-227A as the baseline for primary lithium cells and batteries used to power aircraft equipment. These frameworks directly inform battery selection, containment strategy, and verification scope for ELT and ELT(DT) installations in MEA, and they motivate the broader certification-oriented discussion on energy storage, fire/venting hazards, and maintenance presented in Section 5.

*4.7. MEA-Oriented Design Guidelines and Research Directions*

From the perspective of MEA integration, practical guidelines include: (i) treating ELT/ELT(DT) as a safety-critical load within the onboard power-management hierarchy, (ii) ensuring robust switchover to battery power and budgeting energy across all operating modes, (iii) applying qualification baselines (e.g., DO-160G/AC 21-16G) and incorporating EMC assessment early in the design cycle, and (iv) treating antenna/cabling integrity as a primary safety path in installation design and continued-airworthiness documentation [26,27,32,40].

These considerations motivate research directions aimed at improving energy efficiency under high-rate distress tracking duty cycles, resilience of GNSS and RF front-ends in high-EMI environments, and integration methods compatible with evolving beacon specifications and services in the Cospas–Sarsat ecosystem [16,30,31].

## 5. Certification and Safety Considerations

Certification and safety assurance for ELTs are intrinsically multi-layered, combining equipment-level minimum performance standards with aircraft-level installation and operational constraints. While ELTs must satisfy authority approval baselines and demonstrate compliance with ELT-specific requirements, certification evidence also relies on environmental and electromagnetic qualification and on the safe integration of the energy-storage subsystem that guarantees post-event operation. In more electric aircraft architectures, these issues become more pronounced because converter-dominated power distribution and denser wiring harnesses can alter power quality and the electromagnetic environment experienced by safety-critical equipment. Consequently, compliance is not solely a property of the ELT unit, but also of its installation configuration (antenna/feedline integrity, routing, bonding/grounding, and interface robustness). Moreover, emerging functions such as distress tracking introduce additional operational modes and interfaces that can expand verification scope and increase sensitivity to integration choices. This section therefore reviews the main certification pathways and associated evidence, and discusses how environmental qualification, battery safety, maintenance, and continued airworthiness collectively determine the safety case for ELT/ELT(DT) implementations.

Figure 4 summarizes the layered certification and safety-assurance framework for ELT/ELT(DT) implementations in MEA platforms, showing how operational requirements, beacon interoperability specifications, equipment approval baselines, environmental/EMC qualification, battery safety evidence, and continued airworthiness jointly contribute to the overall safety case.

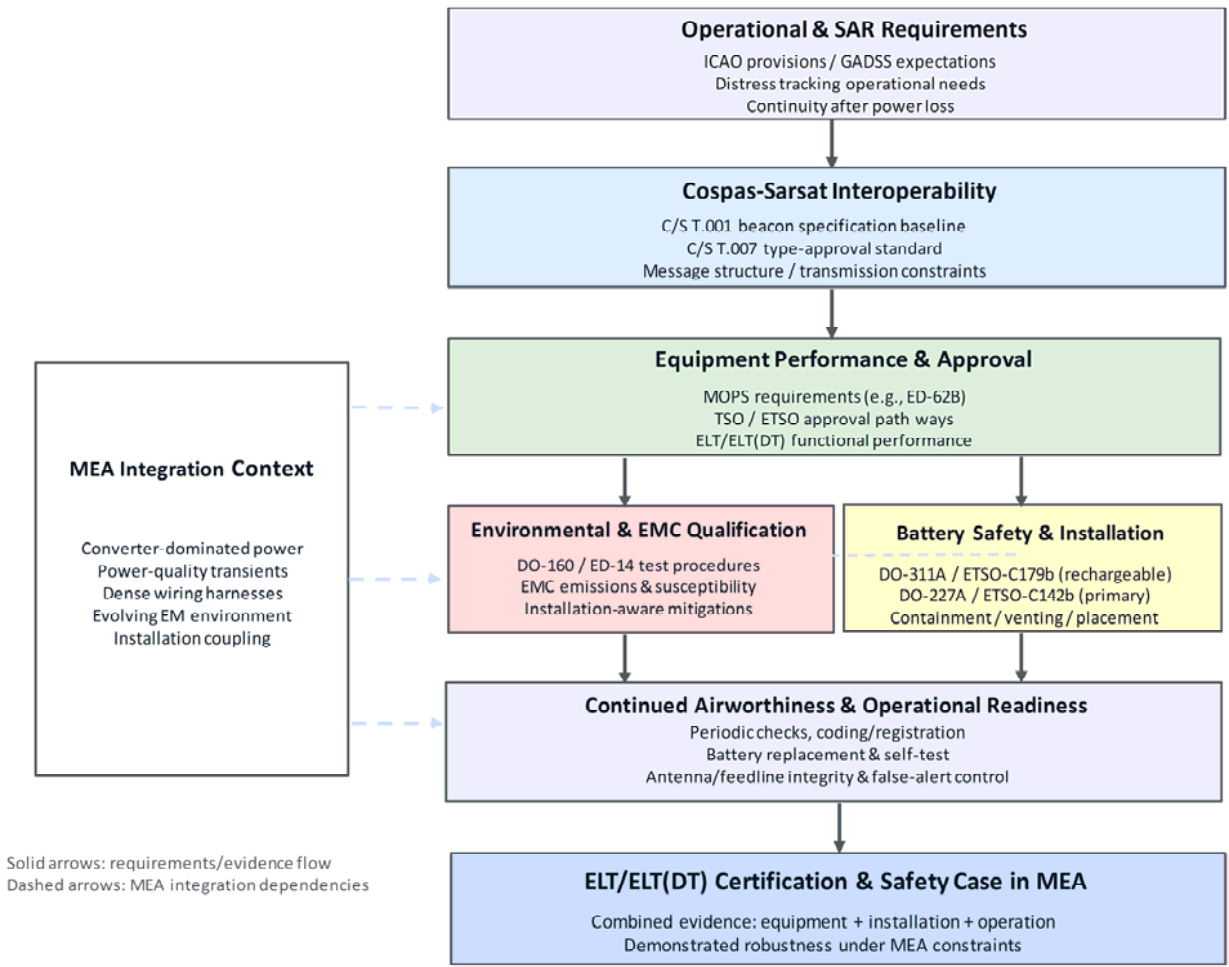

**Figure 4.** Multi-level certification and safety-assurance framework for ELT/ELT(DT) in more electric aircraft (MEA). Operational, qualification, installation, battery, and continued-airworthiness evidence support the overall safety case.

The figure highlights how operational requirements (e.g., GADSS), Cospas-Sarsat interoperability specifications, equipment-level performance and approval baselines (MOPS/TSO/ETSO), environmental and EMC qualification (DO-160/ED-14), battery safety standards and installation constraints, and continued airworthiness activities jointly support certification evidence and the overall safety case.

*5.1. Approval Baselines and Compliance Pathways (TSO/ETSO/MOPS and Cospas–Sarsat)*

Certification and safety assurance for ELTs is built on a layered framework connecting (i) equipment-level performance standards (MOPS), (ii) authority approval mechanisms (TSO/ETSO), and (iii) system-level SAR interoperability requirements governed by the Cospas–Sarsat Programme [4,5,13,16].

For 406 MHz ELTs, authorities reference MOPS/approval baselines such as FAA TSOs and EASA ETSOs. ETSO-C126c explicitly references EUROCAE ED-62B as the minimum performance standard, aligning European approval with a mature baseline for message content, GNSS position handling, crash safety, and installation-related performance [17].

At the system level, interoperability is ensured by Cospas–Sarsat specifications defining beacon transmission characteristics and message structures. C/S T.001 defines the 406 MHz distress beacon specification baseline, while C/S T.007 defines the type approval standard used within the program's approval pathway [16,15].

Operational requirements including GADSS-related obligations for distress tracking are introduced through ICAO provisions. These layers are complementary: TSO/ETSO and MOPS ensure equipment performance, Cospas–Sarsat specifications ensure SAR interoperability, and ICAO provisions define operational expectations that motivate emerging functions such as ELT(DT) [2,29-31].

*5.2. Environmental and EMC Qualification: DO-160/ED-14 and Installation Effects*

Environmental qualification and EMC verification provide evidence that an ELT will operate correctly under physical, electrical, and electromagnetic conditions encountered in service. RTCA DO-160 defines standardized test procedures across environmental categories, and FAA AC 21-16G identifies DO-160 (including DO-160G) as an acceptable means of demonstrating environmental qualification [39,40].

In MEA platforms, qualification must be complemented by installation-aware analysis. Antennas, coaxial feeders, and airframe bonding/grounding networks influence both susceptibility and emissions; thus, compliance is also a function of integration configuration, including wiring routing, separation from power converters and high-current harnesses, and connector retention practices [31,39]. As indicated in Figure 4, DO-160/ED-14 qualification contributes essential certification evidence, but ELT functionality in MEA platforms remains strongly dependent on installation-specific EMC and wiring configuration effects.

Converter-dominated MEA microgrids can amplify conducted-EMI concerns, motivating design-stage EMC assessment rather than late-stage troubleshooting. Simulation-based conducted-EMI analyses can help identify sensitive interfaces and support filter and routing decisions that improve the probability of passing DO-160G categories during qualification [33,34].

DO-160 verification supports certification evidence but does not replace ELT-specific safety objectives: ELTs must still meet ELT MOPS/TSO/ETSO requirements and system-level beacon specification constraints. Integration therefore treats DO-160 qualification as a necessary baseline and uses installation practices as mitigations to maintain ELT functionality under MEA electromagnetic conditions [13,17,16].

*5.3. Battery Safety, Thermal Runaway, and Installation Constraints*

ELTs rely on onboard energy storage to guarantee operation when aircraft power is unavailable, but energy storage introduces hazards that must be controlled through design, test, and installation constraints. RTCA DO-311A defines minimum operational performance standards for rechargeable lithium battery systems, including tests addressing thermal runaway containment; EASA ETSO-C179b references DO-311A as the applicable minimum performance standard for rechargeable lithium systems [33,34]. Figure 4 also emphasizes that battery safety standards and associated ETSO baselines must be interpreted together with installation constraints, since hazard severity depends on placement, containment, venting, and interface design.

For non-rechargeable lithium batteries, RTCA DO-227A provides minimum operational performance standards and EASA ETSO-C142b references DO-227A as the baseline. These standards are directly relevant to ELTs because many installations depend on primary lithium battery packs for long shelf life and low-temperature performance [35,36].

Battery safety is tightly coupled to installation constraints: placement, containment, venting paths, wiring protection, and thermal environment all influence hazard severity. Certification evidence should therefore be consistent with the aircraft-level system safety assessment and with environmental qualification evidence, particularly where batteries interact with charging circuitry or aircraft buses [39,33,35].

Battery considerations also intersect with ELT(DT) functions that may draw aircraft power during in-flight tracking and then revert to battery operation after power loss. This dual sourcing can introduce additional failure modes (e.g., charger faults, bus transients, or unintended depletion) that should be addressed through architecture decisions and verification aligned with both battery and ELT performance baselines [29,30,17].

*5.4. Maintenance, Continued Airworthiness, and Operational Testing*

Continued airworthiness depends on maintenance practices that preserve functional readiness (battery condition, self-test outcomes, correct coding/registration) and installation integrity (antenna system, coaxial routing, connectors, mounting). Operational experience shows that testing procedures evolved with the transition to 406 MHz, requiring appropriate test equipment and procedures beyond legacy practices [14,13]. As summarized in Figure 4, continued airworthiness is part of the ELT/ELT(DT) safety case, linking periodic equipment checks with antenna/feedline integrity, configuration control, and operational readiness.

Full-scale crash tests indicate that cabling and connector integrity can be decisive for post-impact radiated transmission, reinforcing the need for inspection criteria and installation best practices. This supports a maintenance approach that treats antenna/feedline integrity as a primary safety path alongside the ELT unit itself [32].

For ELT(DT) and GADSS-oriented functions, continued airworthiness also includes verification of interfaces used for autonomous activation and GNSS position delivery. Cospas–Sarsat guidance emphasizes that ELT(DT) differs from conventional ELTs in duty cycle and alert distribution, impacting functional checks, configuration control, and inadvertent activation risk [29,30].

Overall, maintenance programmes should connect: (i) equipment-level periodic checks and battery replacement intervals per the approval baseline, (ii) installation inspection focused on antenna and cabling robustness, and (iii) operational procedures for registration, activation control, and post-event alert management to minimize false activations and ensure timely SAR response [2,7,16].

## 6. Emerging Trends and Future Research Opportunities

The next generation of ELTs is shaped by two simultaneous pressures: (i) the modernization of the global SAR ecosystem, which increasingly leverages MEOSAR coverage, multi-constellation GNSS and second-generation beacon services; and (ii) the more electric aircraft paradigm, which tightens constraints on power, thermal integration, wiring density, and electromagnetic compatibility for safety-critical installations [3,19,20,31]. In parallel, ICAO GADSS has crystallized quantitative expectations for tracking and distress information availability (e.g., routine position updates on the order of 15 min and distress-condition updates at the order of 1 min in operational guidance), reinforcing the need for timely, high-integrity position reporting and robust last-known-position (LKP) determination [29–31]. Future ELT developments must preserve the core requirement of autonomous operation while enabling improved diagnostic readiness, richer distress information, and enhanced survivability under crash and fire scenarios, within established approval and type/specification baselines [6,13–17,29–31].

*6.1. Market Readiness and Industry Adoption (ELT Deployments)*

A clear trend is the transition from conventional beacons toward distress tracking variants capable of providing more timely and accurate position reporting and richer operational context. ELT(DT) concepts are intended to support earlier awareness of abnormal situations and to improve the LKP information available to SAR authorities, complementing post-crash beaconing within the broader ICAO GADSS framework [29–31].

Adoption is nevertheless constrained by certification inertia and the need for robust, explainable behavior under faults. Market readiness hinges on demonstrating that added ELT(DT) functions do not compromise the classical ELT role and remain compliant with minimum performance and approval e.g., TSO-C126/ ETSO-2C126, RTCA DO-204A, ETSO-C126c, DO-204B, ED-62B) [4,5,13,17,46, 47]. Accordingly, near-term uptake is expected to favor incremental upgrades—improved multi-constellation GNSS robustness, enhanced self-test and health reporting, and stronger crash-survivability evidence—rather than disruptive architecture changes [6,13,17,31,48]. Recent OEM-level integrations and authorization milestones for ELT(DT) equipment provide encouraging evidence of technical maturity for large transport platforms. Nevertheless, broader adoption remains constrained by aircraft-level integration, certification, validation, and operational implementation efforts [49–51].

*6.2. Energy-Aware Design and Novel Power Sources*

MEA constraints motivate increased attention to energy-aware ELT design, particularly for installations with limited thermal margins and for operators seeking extended maintenance intervals. A key practical driver is that ELT(DT) transmission schedules can be significantly more demanding than classical 406 MHz beaconing. Second-generation specifications define an intensive initial burst phase (24 transmissions at 5 s intervals followed by 18 transmissions at 10 s intervals) and then nominal repetition intervals of 28.5 s after the first 300 s of activation, with randomization around the nominal value [20]. In addition, Cospas–Sarsat specifications define a minimum ELT(DT) operating lifetime of 370 min to maintain the declared power output over the tracking period [16]. Combining the schedule and lifetime implies on the order of ~800 burst transmissions in a single activation event, highlighting why RF/baseband efficiency, GNSS duty-cycling, and low-leakage standby design become central to autonomy assurance [16,20,29,30].

Research directions that are most compatible with certification realities include: (i) lower-power RF and baseband implementations with energy-proportional operation that preserves detectability and message integrity; (ii) battery monitoring and state-of-health (SoH) prognostics aligned with qualification and safety objectives; and (iii) mechanical and thermal packaging approaches that balance energy density with containment, venting/relief, and crash/fire survivability [33–36,39,40,52-54]. Novel power sources (e.g., supercapacitor hybrids or limited energy harvesting) may enable added capabilities or longer stand-by margins, but aviation relevance depends on predictable behavior across environmental extremes and on compliance with established standards for rechargeable and primary lithium systems [33–36,55-58].

*6.3. Advanced SAR Services and Next-Generation Beacons (SGB/RLS/DT)*

Cospas–Sarsat developments toward Second-generation beacons (SGB) establish a roadmap for richer distress messaging and new services within the 406 MHz ecosystem, with Full Operational Capability declared for second-generation ELT(DT) support from 1 January 2024 [59]. From an ELT perspective, SGB enables larger message payloads (e.g., 250-bit second-generation formats) and supports additional fields such as refined location information and beacon state data, thereby increasing the value of robust GNSS input and high-integrity digital burst transmission [20,60,61]. In parallel, MEOSAR brings near-instantaneous global detection and can support rapid location determination (in some cases after a single burst), strengthening the system-level benefit of clean burst detectability and reliable encoded location under adverse installation conditions [19,62].

Return Link Service (RLS) introduces the possibility of acknowledgment to the beacon user in certain contexts. While aviation-specific adoption depends on operational concepts and certification of failure modes, RLS illustrates the broader transition from one-way distress signaling to more interactive SAR workflows [63]. For ELT(DT), the coupling between timely position delivery and SAR performance reinforces the need to treat GNSS input, antenna/coax integrity, and RF behavior as a safety-critical chain, particularly in power-dense MEA environments where installation constraints and EMC threats are elevated [8,13,20,29–31,41,42].

Figure 5 illustrates the future ELT ecosystem within the evolving SAR framework, summarizing the key system interactions between next-generation beacons, MEOSAR services, GNSS inputs, RLS, ground processing, and aircraft integration constraints.

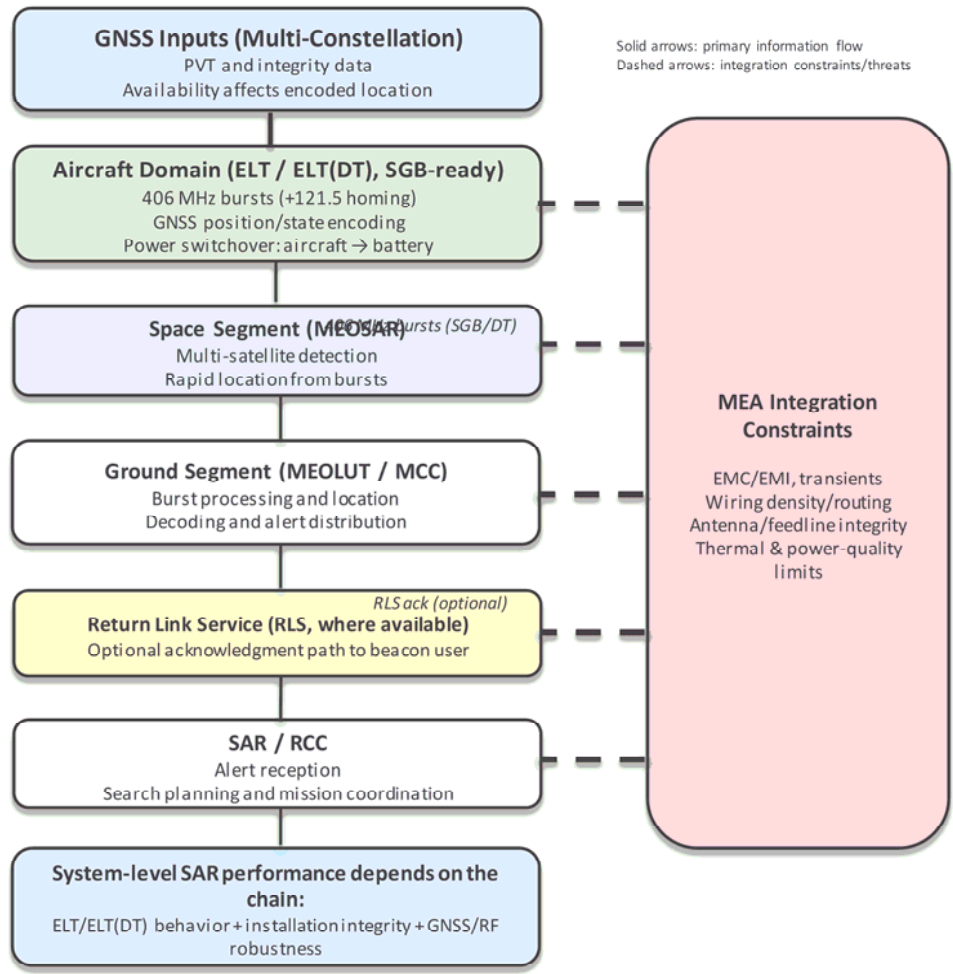

**Figure 5.** Functional overview of next-generation ELT/ELT(DT) integration within the global SAR framework.

*6.4 MEA-driven constraints shaping ELT research*

Beyond beacon-internal advances, ELT and ELT(DT) performance in future aircraft will be increasingly constrained by MEA-wide technological trends.

- Power-architecture migration (HVDC/MVDC/HFAC) and protection: Higher-voltage distribution (±270V/±540V HVDC buses) and faster fault dynamics can constrain ELT installation zoning and wiring segregation, reinforcing the need for robust routing, bonding/grounding, shielding strategy, and connector retention to preserve 406 MHz transmission integrity and GNSS position availability for encoded location and ELT(DT) functions [28, 42, 64-66];

- Power-electronics scale-up and EMI/EMC: wide-bandgap (SiC/GaN) converters and higher switching speeds raise conducted/radiated emissions and common-mode currents; EMC-aware co-design (filters, cable layout, bonding/grounding, shielding terminations, antenna placement) is required to protect both GNSS reception and 406 MHz burst integrity. Experimental GaN-based DC/DC converters demonstrate exceptional performance: >98% efficiency at 100 kHz in UAV applications [67] and 3 kW MEA HVDC interfaces [68]; nevertheless, they require dedicated EMI mitigation as conducted emissions penetrate ELT GNSS front-ends, with multi-source dual active bridge (DAB) topologies amplifying susceptibility risks [69];

- Thermal management and integration density: higher power density in equipment bays reduces thermal margins and installation flexibility, impacting battery endurance, self-test reliability, and survivability-driven placement decisions for ELT units and antennas in MEA platforms [70-72].

- Energy storage qualification and safety envelope: advanced batteries and hybrid storage can support richer functionality, but certification constraints (thermal-runaway containment, predictable aging, maintenance policies) will continue to shape feasible ELT/ELT(DT) power strategies and verification scope [74]. Recent analyses establish Thermal Runaway Explosion Index metrics for aviation lithium systems [75], while equipment-level testing validates non-rechargeable ELT battery containment under DO-227A [35] and NASA defines aging models ensuring predictable performance across certification envelopes [76].
- Prognostic and health management (PHM), digitalization, and cyber-physical assurance: PHM, together with broader digitalization trends, strengthens the case for considering GNSS interfacing and ELT installation integrity as safety-critical paths, especially as health monitoring and connected maintenance become more pervasive [77]. At the hardware level, representative mixed-signal research, such as time-interleaved sigma-delta digital-to-analog converters, illustrates the need for robust interfaces at the signal-chain boundary, thereby complementing data-interface assurance at higher abstraction levels 78,79]. Furthermore, increasing system connectivity motivates safety-security co-assurance for any ELT(DT) functions relying on data interfaces or embedded diagnostics, since such functions may also be exposed to cybersecurity threats [80].

*6.5. Research Gaps and Open Challenges*

Across the previous sections, several open issues emerge that cut across technology, integration, and certification. The most relevant gaps and challenges are:

- Survivability-driven design evidence. Crash studies show that failures can arise from antenna damage, mounting failure, or post-impact shielding, indicating that installation and mechanical protection are often as critical as transmitter compliance. There is a clear need for standardized, installation-aware survivability metrics and repeatable test methods applicable to modern architectures, including ELT(DT) [6,32].
- EMC robustness under power-electronics-rich environments. MEA power architectures increase EMI risk and constrain installation routing options. There is a continuing need for validated mitigation strategies (routing, shielding, bonding/grounding, connector retention) that preserve both 406 MHz transmission integrity and GNSS performance across installation zones, and for practical guidance linking qualification results to aircraft-level integration decisions [40,42].
- Energy autonomy versus richer functionality. Features such as more frequent position updates, extended monitoring, or additional communication links can increase energy demand. The key challenge is enabling improved functionality without eroding the endurance margins required by standards and without introducing new failure modes that complicate certification [11,13–15,27–30].
- The key challenge is enabling enhanced functionality without eroding endurance margins required by standards and without introducing new failure modes that complicate certification. In practice, near-term opportunities align with battery/power-management within DO-311A (rechargeable) and DO-227A (primary) envelopes while remaining consistent with beacon-level specifications (C/S T.001; SGB in T.018 where applicable) [33,35,16].
- Certification pathways for evolving SAR services. As the SAR ecosystem evolves (MEOSAR, SGB and related services), certification frameworks must balance stability with innovation. Translating system-level benefits into certifiable equipment requirements and demonstrating safety under abnormal and fault conditions remain a core challenge, particularly for functions that blur the boundary between "distress beacon" and "tracking/monitoring system" [16-19].

## 7. Conclusions

This review has examined the evolution of emergency locator transmitters through the dual lens of (i) the modernization of the satellite-based search-and-rescue ecosystem and (ii) the progressive electrification of aircraft systems under the MEA paradigm. The historical transition from legacy 121.5/243 MHz beacons to digitally encoded 406 MHz ELTs while keeping 121.5 MHz mainly as a homing signal in combined 406/121.5 units marked a step change in global detectability, identification, and localization performance, especially after the end of satellite processing of 121.5/243 MHz alerts in 2009. Subsequent integration of GNSS and MEOSAR further reduced detection latency and strengthened the value of clean burst detectability and high-integrity encoded position.

From a MEA perspective, the main conclusion is that ELT performance and compliance can no longer be interpreted as a transmitter-only property. Instead, it emerges from a coupled 'equipment + installation + aircraft electrical environment' system. Converter-dominated power architectures, denser harnessing, and tighter equipment bays increase sensitivity to power quality disturbances, thermal constraints, and EMI/EMC interactions, pushing ELT integration toward a design-for-robustness approach rather than a late-stage qualification exercise. In practice, this means that aircraft-level constraints (segregation, routing, bonding/grounding, shielding termination quality, connector retention, and zonal placement) become as influential as nominal RF output or message format compliance.

Another key finding is that installation survivability is a recurring limiting factor in real-world effectiveness. Full-scale crash tests and survivability studies show that loss of radiated transmission can be due by antenna damage, mount failure, connector disengagement, feedline discontinuities, or post-impact shadowing, even when the ELT unit itself is nominally compliant. This supports the need for standardized, installation-aware survivability metrics and repeatable assessment methods applicable to modern aircraft and to emerging ELT(DT) architectures.

Third, the energy subsystem remains the dominant enabler and, increasingly, a dominant constraint when ELTs are considered in MEA settings. Classical ELT requirements already mandate independent energy storage and endurance across harsh environmental envelopes; however, distress tracking and second-generation behaviors can significantly increase burst counts and total energy demand relative to traditional post-crash alerting. Consequently, endurance assurance must be treated as an energy budgeting problem across modes (standby, self-test, GNSS acquisition, tracking, post-crash alerting), not only as a "battery capacity" issue. Battery selection and integration are also bounded by certification frameworks: rechargeable systems should comply with DO-311A/ETSO-C179b, while primary lithium systems must comply with DO-227A/ETSO-C142b, shaping containment, ventilation strategy, installation restrictions, and verification scope for ELT/ELT(DT) in MEA platforms.

Fourth, certification evidence for MEA-integrated ELTs is inherently multilayered. Compliance should reconcile (i) ELT-specific performance baselines (MOPS/TSO/ETSO), (ii) Cospas–Sarsat beacon specifications and evolving service expectations (including SGB-related provisions where applicable), and (iii) environmental/EMC qualification (e.g., DO-160G accepted via AC 21-16G), while also accounting for aircraft-level installation and continued-airworthiness practices. For ELT(DT), the addition of in-flight operational modes and interfaces expands the safety case: the design must demonstrate that new tracking features do not compromise the classical distress function and that fault behavior remains deterministic and explainable under abnormal electrical and electromagnetic conditions.

Finally, the review highlights a set of research directions that are most likely to produce certifiable progress for MEA-optimized ELTs: (i) survivability-driven installation engineering with better standardized metrics; (ii) energy-aware architectures that improve RF/baseband efficiency and GNSS duty cycling without eroding mandated endurance; (iii) EMC-aware co-design linking early prediction methods to installation mitigations and DO-160 outcomes; and (iv) integration concepts for next-generation SAR services (MEOSAR/SGB and, where relevant, RLS) that preserve the 'critical chain' from GNSS input to antenna/feedline integrity to 406 MHz burst transmission.

In general, ELTs remain a mature and highly regulated safety technology, but MEA electrification and SAR modernization are shifting the engineering center of gravity toward integration: the most impactful advances will come from solutions that reduce energy demand, preserve the EMC margin, and strengthen installation survivability, while remaining compatible with established certification pathways and the evolving Cospas–Sarsat ecosystem.

**Funding:** This research received no external funding.